\documentclass[10pt,conference]{IEEEtran}

\usepackage{cite}
\usepackage{amsmath,amssymb,amsfonts}
\usepackage{algorithmic}
\usepackage{graphicx}
\usepackage{textcomp}
\usepackage{xcolor}
\usepackage{booktabs}
\usepackage{multirow}
\usepackage{listings}
\usepackage{url}
\usepackage{hyperref}
\usepackage{tikz}
\usepackage{pgfplots}
\pgfplotsset{compat=1.18}
\usetikzlibrary{arrows.meta, positioning, calc}

\begin{document}

\title{AgentModernize: Preserving Business Logic in Legacy Modernization with Multi-Agent LLMs and Behavioral Specification Graphs}

\author{%
\IEEEauthorblockN{Sheikh Nazib Ahmed}
\IEEEauthorblockA{University of Texas at Arlington\\
Arlington, TX, USA\\
sxa5256@mavs.uta.edu}
\and
\IEEEauthorblockN{Marnim Galib}
\IEEEauthorblockA{University of Texas at Arlington\\
Arlington, TX, USA\\
marnim.galib@uta.edu}
}

\maketitle

\begin{abstract}
Legacy modernization breaks business logic more often than most teams expect. Most tools and LLM-based approaches treat modernization as syntax translation: convert COBOL to Java, swap PL/SQL for Python, ship it. Implicit rules, edge-case handling, and cross-module constraints that keep production systems running are lost, and nobody notices until something fails in production.

We present \textit{AgentModernize}, a multi-agent framework that treats modernization as a behavioral preservation problem. Four agents handle extraction, specification, code generation, and validation. The key intermediate artifact, a \textit{Behavioral Specification Graph (BSG)}, forces extracted business logic to be explicit and inspectable before any code is generated.

We evaluated on LegacyModernize-8, eight synthetic scenarios spanning telecom and banking, under a fair protocol where each method's tests are generated from its own API surface (3 trials, temperature 0.0). With GPT-4o-mini, AgentModernize with feedback achieves 23.0\% mean BER (non-zero on 5/8 scenarios, up to 53.3\%), while AgentModernize without feedback reaches 23.8\%. SP-LLM scores 12.4\% (non-zero on 2/8) and CoT-LLM 4.5\% (2/8). No single method dominates all scenarios. The feedback loop is decisive for scenarios requiring iterative correction (S4, S6, S7) but can regress code in others (S2, S8). The BSG captures 92.3\% of gold-standard rules with 90.2\% precision; the bottleneck is code generation, not extraction. A cross-model study (GPT-4o, GPT-5.3-codex) reveals the pipeline's benefit is inversely correlated with model capability: stronger models achieve higher BER with single-prompt methods than with the pipeline. For regulated industries, the pipeline's traceable artifacts (business rule inventory, BSG, equivalence reports) provide an audit trail that no single-prompt approach can match.
\end{abstract}


%% ============================================================
\section{Introduction}

Telecommunications carriers, financial institutions, and healthcare providers run their core operations on systems built twenty or thirty years ago. COBOL batch jobs process billing. PL/SQL procedures enforce contract logic. Shell scripts orchestrate provisioning workflows. These systems work, and the business rules encoded in them have been refined through years of production experience, bug fixes, and regulatory adaptation~\cite{comella2000survey}. They are also expensive to maintain, difficult to integrate with modern platforms, and increasingly fragile as the engineers who wrote them retire~\cite{breivold2012systematic}.

Modernization is the standard response, but most modernization efforts destroy the thing they should preserve: behavioral semantics. Business rules in legacy systems are rarely documented; they live in control flow patterns, conditional branches, exception handlers, and configuration files~\cite{kazman1998requirements}. A line-by-line COBOL-to-Java translation can compile and run yet silently break edge-case handling, alter validation logic, or drop constraints that were never written down~\cite{rugaber2004model}. We saw this firsthand in a telecom provisioning workflow where a syntactically correct translation passed compilation but silently dropped a suspended-account exemption that had been in production for fifteen years. The system looked modern and behaved differently.

Large Language Models can parse legacy code, reason about intent, and generate modern equivalents~\cite{chen2021evaluating, li2023starcoder}, but prompting one to ``convert this legacy code to a modern API'' is brittle. Context windows cannot hold an entire legacy codebase; a single generation pass provides no way to verify behavior was preserved; and a monolithic prompt conflates understanding, specification, transformation, and validation, tasks that benefit from separation~\cite{hou2024llm4se}.

Our approach, \textit{AgentModernize}, decomposes legacy modernization into four agent-handled phases:

\begin{enumerate}
    \item A \textbf{Legacy Analyzer} that extracts both explicit and implicit business rules, along with control flows and operational constraints, from legacy artifacts.
    \item A \textbf{Specification Generator} that encodes the extracted logic into \textit{Behavioral Specification Graphs (BSGs)}, a structured intermediate representation with preconditions, postconditions, and invariants.
    \item A \textbf{Modernization Transformer} that generates modern service implementations from the BSG, preserving behavioral contracts.
    \item An \textbf{Equivalence Validator} that checks the generated code against the BSG through automated test generation and differential trace analysis, feeding failures back for iterative correction.
\end{enumerate}

The BSG acts as a ``glass box'' between legacy understanding and modern generation: extracted business logic is explicit and inspectable before any code is written. When the Validator detects a behavioral divergence, it triggers a targeted correction loop rather than a full re-generation. As our ablation shows (Table~\ref{tab:rq4}), the multi-agent decomposition consistently outperforms single-prompt baselines, and the feedback loop adds value on scenarios with localized, correctable errors.

We evaluate on LegacyModernize-8, a benchmark of eight synthetic scenarios covering telecom and banking modernization, measuring behavioral equivalence rate and business rule preservation against single-prompt and chain-of-thought LLM baselines. With GPT-4o-mini, AgentModernize achieves 23.0\% mean BER with feedback and 23.8\% without, compared to 12.4\% for SP-LLM and 4.5\% for CoT-LLM. No method dominates all scenarios; the feedback loop helps on some and regresses others, a nuance prior single-pass evaluations would miss.

\subsection{Contributions}

Our contributions are:

\begin{itemize}
    \item A \textbf{multi-agent framework} that decomposes legacy modernization into extraction, specification, transformation, and validation, with a feedback loop for iterative correction.
    \item \textbf{Behavioral Specification Graphs}, an intermediate representation that captures business rules, pre/post-conditions, data constraints, and control flow dependencies in an inspectable, verifiable form.
    \item An \textbf{automated equivalence checking approach} based on test oracle generation from BSG specifications, combined with differential trace analysis.
    \item An \textbf{empirical evaluation} on an eight-scenario benchmark (including a held-out banking scenario) with ablation and model comparison, showing that the multi-agent pipeline with feedback achieves the highest BER on scenarios requiring iterative correction, while revealing that feedback can regress code in other scenarios.
\end{itemize}

%% ============================================================
\section{Related Work}

\subsection{Legacy System Modernization}

Early frameworks catalogued modernization strategies ranging from wrapping to full re-engineering~\cite{comella2000survey} and formalized the process as iterative abstraction, transformation, and refinement (the horseshoe model~\cite{kazman1998requirements}). Model-driven approaches like MoDisco~\cite{bruneliere2014modisco} formalize legacy knowledge using KDM and ADM standards, but constructing the models requires manual effort that often rivals the cost of the modernization itself.

Industry tools (IBM Rational Asset Analyzer, Micro Focus Enterprise Analyzer, COBOL-to-Java transpilers) handle syntactic translation competently~\cite{delucia2001cobol}, but ``compile-and-run'' is not behavioral equivalence. None of these tools verify that the translated system \textit{behaves} the same as the original, and none publish behavioral equivalence metrics we could compare against. As Section~\ref{sec:fair-eval} shows, even LLM-based approaches fail to preserve behavior without an explicit feedback mechanism.

\textbf{Design/Workflow Recovery and Code Summarization.} Adjacent literatures on program comprehension have long addressed pieces of the modernization problem under different names. Zou and Hung~\cite{zou2006workflow} extract workflow structure from e-commerce applications; the recent survey by Zhang et al.~\cite{zhang2024summarization} covers automatic source code summarization techniques that produce natural-language descriptions of code behavior. Our Legacy Analyzer is closest in spirit to workflow recovery, and the BSG plays a role analogous to a structured code-level summary. What separates AgentModernize from this line of work is the closed loop with a code-generation and verification stage: recovery/summarization systems traditionally stop at the human-readable artifact, whereas the BSG must be executable-enough for downstream synthesis and differential testing.

\textbf{Comparison to Established IRs.} The Knowledge Discovery Metamodel (KDM)~\cite{omg2016kdm} and Architecture-Driven Modernization (ADM) standards from OMG provide formal representations of legacy system semantics. Behavioral interface specification languages such as JML~\cite{leavens1999jml} and Eiffel~\cite{meyer1992dbc} encode pre/post/invariants per method. BSG draws from both but differs in three respects: it is designed for LLM-mediated extraction with per-rule confidence scoring and source traceability; it distinguishes explicit from implicit rules (a categorization absent from KDM/JML); and it serves as an inspectable trust boundary between LLM extraction and LLM generation, which had no analog in pre-LLM workflows. BSG adapts the graph-IR-with-contracts paradigm to LLM-driven modernization rather than introducing a new representation.

\subsection{LLMs for Software Engineering}

LLMs have shown strong results in code generation~\cite{chen2021evaluating}, bug detection~\cite{li2023llmanalysis}, code review~\cite{li2022codereviewer}, and program repair~\cite{xia2022repair}. The work most relevant to ours is Pan et al.~\cite{pan2024lost}, who found that GPT-4 produces syntactically correct cross-language translations but introduces semantic bugs in complex control flows. This is consistent with our own experiments (Section~\ref{sec:fair-eval}): single-prompt baselines compiled and ran but achieved low BER (12.4\% SP-LLM, 4.5\% CoT-LLM mean) compared to the multi-agent pipeline.

The common thread in all of these studies is that they treat modernization as a single-pass task. One model, one prompt, one output. No intermediate representation, no verification, no feedback. For a self-contained utility function, that might work. For a 280-line COBOL provisioning workflow with thirteen interacting business rules and implicit logic buried in exception handlers, it does not. Our results in Table~\ref{tab:rq1} confirm this empirically.

\subsection{Multi-Agent LLM Systems}

MetaGPT~\cite{hong2024metagpt} partitions software development across architect, engineer, and tester agents. ChatDev~\cite{qian2024chatdev} models an entire software company through multi-agent dialogue. AgentCoder~\cite{huang2023agentcoder} pairs a code generator with an adversarial test generator. Self-Refine~\cite{madaan2023selfrefine} and Reflexion~\cite{shinn2023reflexion} show that LLM agents can improve their own outputs through self-generated feedback, a principle our feedback loop shares.

All of these target \textit{greenfield} development. Building new code from scratch is a different problem than modernizing existing code with undocumented business rules; none extract implicit logic from decades-old COBOL, represent it in a verifiable form, or confirm that generated code preserves undocumented behavior. AgentModernize's BSG is task-specific to modernization (not applicable to code review or refactoring in general), but it is not tied to any application domain: it captures business rules, control/data flow, and constraints in a form that applies equally to telecom, banking, healthcare, or government legacy systems, as our S8 result begins to show.

\subsection{Behavioral Equivalence}

Formal program equivalence (bisimulation, trace equivalence, observational equivalence) provides the theoretical grounding for our verification approach~\cite{milner1989communication}. Full formal verification is intractable for most real-world systems. Practitioners rely on lighter alternatives: differential testing~\cite{mckeeman1998differential} compares outputs across system versions; metamorphic testing~\cite{chen1998metamorphic} checks whether known input transformations produce expected output transformations.

Our Equivalence Validator combines both ideas but generates test oracles from the BSG specification, not from the legacy code directly. This avoids the circularity of testing a translation against itself. We do not claim formal guarantees; Section~\ref{sec:fair-eval} shows the approach catches a class of behavioral regressions that single-pass methods miss.

%% ============================================================
\section{The AgentModernize Framework}

\subsection{Overview}

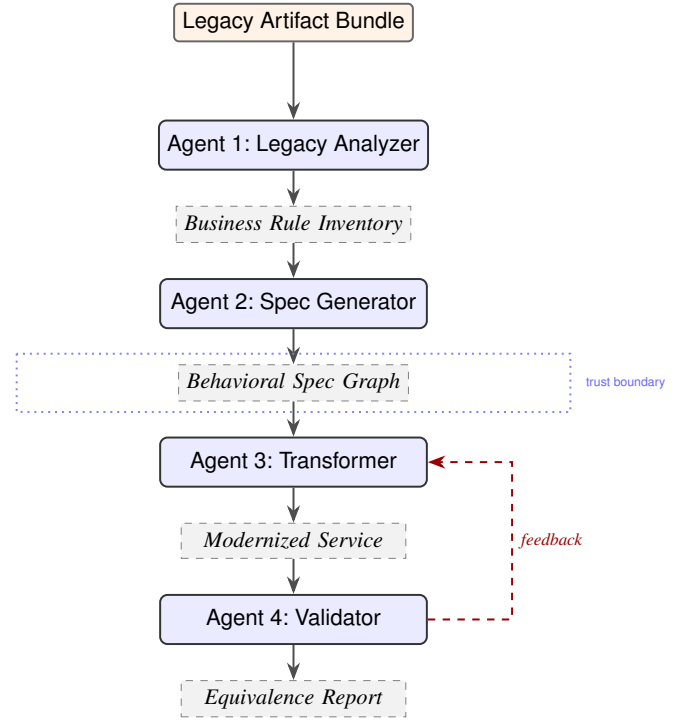
\begin{figure}[htbp]
\centering
\resizebox{0.48\textwidth}{!}{%
\begin{tikzpicture}[
  node distance=1.1cm and 0.5cm,
  agent/.style={rectangle, draw=black!80, fill=blue!8, thick, rounded corners=3pt,
                minimum width=3.8cm, minimum height=0.7cm, font=\small\sffamily},
  artifact/.style={rectangle, draw=black!50, fill=gray!10, dashed,
                   minimum width=3.2cm, minimum height=0.5cm, font=\small\itshape},
  input/.style={rectangle, draw=black!60, fill=orange!10, thick, rounded corners=2pt,
                minimum width=3.2cm, minimum height=0.5cm, font=\small\sffamily},
  arr/.style={-{Stealth[length=2.5mm]}, thick, black!70},
  farr/.style={-{Stealth[length=2.5mm]}, thick, red!60!black, dashed},
]
% Input
\node[input] (input) {Legacy Artifact Bundle};

% Agent 1
\node[agent, below=of input] (a1) {Agent 1: Legacy Analyzer};
\node[artifact, below=0.5cm of a1] (bri) {Business Rule Inventory};

% Agent 2
\node[agent, below=0.5cm of bri] (a2) {Agent 2: Spec Generator};
\node[artifact, below=0.5cm of a2] (bsg) {Behavioral Spec Graph};

% Agent 3
\node[agent, below=0.5cm of bsg] (a3) {Agent 3: Transformer};
\node[artifact, below=0.5cm of a3] (code) {Modernized Service};

% Agent 4
\node[agent, below=0.5cm of code] (a4) {Agent 4: Validator};
\node[artifact, below=0.5cm of a4] (report) {Equivalence Report};

% Forward arrows
\draw[arr] (input) -- (a1);
\draw[arr] (a1) -- (bri);
\draw[arr] (bri) -- (a2);
\draw[arr] (a2) -- (bsg);
\draw[arr] (bsg) -- (a3);
\draw[arr] (a3) -- (code);
\draw[arr] (code) -- (a4);
\draw[arr] (a4) -- (report);

% Feedback loop
\draw[farr] (a4.east) -- ++(1.2,0) |- node[right, pos=0.25, font=\scriptsize\itshape, text=red!60!black] {feedback} (a3.east);

% Trust boundary annotation
\draw[thick, dotted, blue!50] ([xshift=-2.3cm, yshift=0.15cm]bsg.north west) rectangle ([xshift=2.3cm, yshift=-0.15cm]bsg.south east);
\node[font=\tiny\sffamily, blue!60, anchor=west] at ([xshift=2.4cm]bsg.east) {trust boundary};

\end{tikzpicture}%
}
\caption{AgentModernize Pipeline Architecture. Solid arrows show the forward pipeline; the dashed red arrow shows the feedback loop from the Validator to the Transformer. The BSG (dotted box) acts as a trust boundary between extraction and generation.}
\label{fig:architecture}
\end{figure}

The framework takes a bundle of legacy artifacts (source code, configuration files, database schemas, and any available documentation) and produces a modernized service implementation with a report on how faithfully the original behavior was preserved. Four agents handle this in sequence, each reading from and writing to a shared pipeline state. Fig.~\ref{fig:architecture} shows the architecture.

The pipeline is not purely feed-forward: when the Equivalence Validator (Agent~4) detects behavioral divergences, it feeds specific failure information back to the Modernization Transformer (Agent~3) for targeted correction. This loop runs up to three iterations.

\textbf{LLM Interaction Model.} All four agents use zero-shot prompting with off-the-shelf models via their public APIs. We do not fine-tune, do not use RAG, and do not include few-shot in-context examples. Each agent's prompt is a fixed instruction template plus the current pipeline state. The scenarios in Section~\ref{sec:benchmark} were used during development to iterate on prompt templates, but no model weights are updated and no scenario-specific examples are embedded in prompts at inference time.

\subsection{Legacy Artifact Bundle}

The input to the framework is a \textit{Legacy Artifact Bundle} $L = \{S, C, D, N\}$ where:

\begin{itemize}
    \item $S$ = Source code files (e.g., COBOL programs, PL/SQL procedures, shell scripts)
    \item $C$ = Configuration files (e.g., JCL, parameter files, cron schedules)
    \item $D$ = Database schemas (e.g., DDL statements, copybook layouts, data dictionaries)
    \item $N$ = Natural language documentation (optional; e.g., runbooks, comments, requirement documents)
\end{itemize}

\subsection{Agent 1: Legacy Analyzer}

The Legacy Analyzer extracts both explicit business rules (stated directly in conditionals, guards, or comments) and implicit rules (inferable from code patterns, default values, exception handlers, or cross-module dependencies). Each extracted rule is tagged with its category (\texttt{explicit} or \texttt{implicit}) and a confidence score in the Business Rule Inventory. The agent operates in three sub-phases:

\textbf{Phase 1a --- Structural Parsing:} The agent identifies program entry points, module boundaries, data flows, and external dependencies. For procedural legacy code, this involves identifying PERFORM/CALL hierarchies, file I/O operations, and database interactions.

\textbf{Phase 1b --- Business Rule Extraction:} The agent identifies conditional logic, validation rules, computation formulas, state transitions, and exception handling patterns. Each extracted rule is annotated with: a natural language description, the source location (file, line range), input variables and output effects, and a confidence score (high/medium/low) based on extraction clarity.

\textbf{Phase 1c --- Constraint Discovery:} The agent identifies implicit constraints including data type restrictions, value ranges, referential integrity rules, temporal ordering requirements, and business-specific invariants (e.g., ``order total must equal sum of line items'').

\textbf{Output:} A \textit{Business Rule Inventory (BRI)}, a JSON document containing all extracted rules, constraints, and metadata.

\subsection{Agent 2: Specification Generator}

The Specification Generator transforms the Business Rule Inventory into a \textit{Behavioral Specification Graph (BSG)}, the core intermediate representation of AgentModernize.

\textbf{Definition 1 (Behavioral Specification Graph).} A BSG is a directed acyclic graph $G = (V, E, Pre, Post, Inv)$ where:

\begin{itemize}
    \item $V$ is a set of \textit{operation nodes}, each representing a discrete business operation (e.g., ``validate order'', ``calculate tax'').
    \item $E \subseteq V \times V \times L$ is a set of \textit{labeled edges} representing control flow, where $L = \{sequence, conditional, parallel, error\}$ is the label set. Data dependencies live at the node level through the $inputs(v)$ and $outputs(v)$ annotations below; a data dependency from $u$ to $v$ is implicit when $outputs(u) \cap inputs(v) \neq \emptyset$. Keeping control flow on edges and data flow on nodes lets downstream agents and human reviewers inspect each independently.
    \item $Pre: V \rightarrow \mathcal{P}$ maps each node to a set of \textit{preconditions}, logical predicates that must hold before the operation executes.
    \item $Post: V \rightarrow \mathcal{P}$ maps each node to a set of \textit{postconditions}, predicates that must hold after it completes.
    \item $Inv: G \rightarrow \mathcal{P}$ is a set of \textit{global invariants} that must hold throughout the workflow.
\end{itemize}

Each operation node $v \in V$ is further annotated with: $inputs(v)$ (required input data fields with types), $outputs(v)$ (produced output data fields with types), $rules(v)$ (set of business rule IDs from the BRI), and $error\_behavior(v)$ (expected behavior under error conditions).

\textbf{Edge Label Taxonomy.} The four edge labels map to distinct code-generation patterns: \texttt{sequence} is unconditional next-step ordering ($v$ always executes after $u$, generated as a straight-line call); \texttt{conditional} is a guarded transition where the guard is stored as a $Pre(v)$ clause (typically paired with a sibling \texttt{conditional} or \texttt{error} edge); \texttt{parallel} means $u$ and $v$ have no data dependency and can run concurrently (generated as an async/gather block); \texttt{error} is an exception-triggered transition matched against $error\_behavior(u)$ (generated as an except/catch handler). Every edge carries exactly one label; patterns outside this set (e.g., cooperative coroutines) are out of scope for the current BSG.

\textbf{Concrete Example (S1, Order Validation).} The BSG node below, extracted from S1's COBOL order validation program, illustrates the representation:

\begin{lstlisting}[caption={BSG Node --- S1 Disconnect Order Validation}, language={}]
{
  "operation": "ValidateDisconnectOrder",
  "source_rule": "BR-004",
  "source_location": "ORDER_VALIDATION.cob:118-142",
  "preconditions": [
    "account_status in ['ACTIVE', 'SUSPENDED']",
    "order_type == 'DISCONNECT'"
  ],
  "postconditions": [
    "suspended accounts proceed for disconnect orders",
    "non-disconnect orders rejected for suspended"
  ],
  "invariants": ["order_total == sum(line_items)"],
  "confidence": "high"
}
\end{lstlisting}

This node captures an implicit rule (suspended accounts are normally blocked, but disconnect orders are exempted) that would be lost in a direct syntax translation. The BSG makes it explicit before code generation begins.

\subsection{Agent 3: Modernization Transformer}

The Modernization Transformer generates a modern service-oriented implementation from the BSG. It operates under four transformation principles:

\textbf{Principle 1 --- Operation-to-Endpoint Mapping:} Each BSG operation node maps to a discrete function or API endpoint in the modernized system.

\textbf{Principle 2 --- Contract Preservation:} Pre/postconditions from the BSG are translated into input validation, output assertions, and error handling in the generated code.

\textbf{Principle 3 --- Data Model Derivation:} Input/output annotations from BSG nodes are aggregated to derive modern data models (e.g., Pydantic models, dataclasses) with type constraints.

\textbf{Principle 4 --- Flow Preservation:} BSG edge labels (sequence, conditional, parallel, error) determine the orchestration pattern in the modernized implementation.

\textbf{Output:} A \textit{Modernized Service Package} with API endpoint implementations (Python/FastAPI), data models with validation rules, orchestration logic preserving the BSG control flow, error handling matching the BSG error behaviors, and documentation mapping each endpoint to its source BSG node.

\subsection{Agent 4: Equivalence Validator}

The Equivalence Validator verifies that the modernized implementation matches the behavioral contract specified by the BSG, using three validation types:

\textbf{Type 1 --- Contract Verification:} For each BSG node, the agent generates test cases that exercise the corresponding modernized endpoint, verifying that preconditions are enforced, postconditions hold, and invariants are maintained across multi-step workflows.

\textbf{Type 2 --- Boundary Testing:} The agent generates boundary test cases based on BSG constraints (e.g., minimum/maximum values, empty inputs, null handling, type boundaries).

\textbf{Type 3 --- Differential Trace Analysis:} The agent constructs end-to-end workflow traces from the BSG and executes them against the modernized implementation, comparing actual outputs against expected BSG-specified outputs.

\textbf{Output:} A \textit{Behavioral Equivalence Report (BER)} containing: per-node equivalence status (PASS/FAIL/PARTIAL), overall behavioral equivalence rate, failed test cases with root cause analysis, and recommendations for remediation.

\textbf{Feedback Loop:} When the BER identifies behavioral divergences, the pipeline re-invokes the Modernization Transformer with targeted correction instructions, including the failing test cases and expected behaviors. This feedback loop runs for a configurable maximum number of iterations (default: 3).

\subsection{Orchestration and State Management}

The four agents are orchestrated by a directed-graph execution engine (LangGraph). A shared state object accumulates artifacts across phases:

\begin{lstlisting}[caption={Pipeline State Object}]
State = {
  legacy_bundle: LegacyArtifactBundle,
  business_rules: BusinessRuleInventory,
  bsg: BehavioralSpecificationGraph,
  modern_code: ModernizedServicePackage,
  equiv_report: BehavioralEquivalenceReport,
  iteration: int,
  status: RUNNING | COMPLETED | FAILED
}
\end{lstlisting}

Each agent reads from and writes to this state, giving full traceability from legacy artifacts to modernized output.

%% ============================================================
\section{Evaluation Design}

\subsection{Research Questions}

We evaluate AgentModernize with respect to the following research questions:

\begin{itemize}
    \item \textbf{RQ1 (Behavioral Equivalence):} To what extent does AgentModernize preserve the behavioral semantics of legacy systems in modernized implementations?
    \item \textbf{RQ2 (Business Rule Preservation):} How effectively does the framework extract and preserve individual business rules through the modernization pipeline?
    \item \textbf{RQ3 (Effort Reduction):} How does the manual effort required for AgentModernize-assisted modernization compare to single-prompt LLM baselines?
    \item \textbf{RQ4 (Agent Contribution):} What is the individual contribution of each agent to the overall modernization quality?
    \item \textbf{RQ5 (BSG Quality):} How accurately does the Behavioral Specification Graph capture the gold-standard business rules?
    \item \textbf{RQ6 (Model Sensitivity):} How does modernization quality vary across LLM model sizes?
\end{itemize}

\subsection{Benchmark: LegacyModernize-8}
\label{sec:benchmark}

\textbf{All eight scenarios in LegacyModernize-8 are synthetic}, hand-authored to reflect realistic legacy patterns (COBOL/PL-SQL procedures with embedded business logic, multi-step workflows, and implicit rules in exception handlers and default values) rather than sampled from production codebases. Seven scenarios are drawn from telecommunications and one from banking (S8) to test generalization beyond telecom. Each scenario is a legacy artifact bundle plus a gold-standard behavioral specification curated by a domain expert (AI-assisted drafting, human-reviewed). Table~\ref{tab:benchmark} summarizes them.

\begin{table}[htbp]
\caption{LegacyModernize-8 Benchmark Scenarios}
\label{tab:benchmark}
\centering
\footnotesize
\begin{tabular}{@{}clllcc@{}}
\toprule
\textbf{ID} & \textbf{Scenario} & \textbf{Domain} & \textbf{Complexity} & \textbf{Rules} & \textbf{LOC} \\
\midrule
S1 & Order Validation & Orders & Medium & 12 & 248 \\
S2 & Billing Disputes & Billing & High & 12 & 195 \\
S3 & Service Activation & Provisioning & High & 14 & 245 \\
S4 & Circuit Inventory & Inventory & High & 13 & 280 \\
S5 & Fault Escalation & Fault Mgmt & High & 13 & 230 \\
S6 & Contract Renewal & Contracts & High & 12 & 210 \\
S7 & Account Migration & Accounts & Very High & 15 & 280 \\
S8 & Bank Transactions & Banking & Very High & 15 & 310 \\
\bottomrule
\end{tabular}
\end{table}

\textbf{Scenario Design.} Each scenario contains explicitly documented rules and at least three implicit rules inferable only from code patterns. Complexity reflects the number of conditional branches, external dependencies, and state transitions. Scenarios S1--S7 cover telecom-domain challenges: multi-step validation (S1), tier-based conditional logic (S2), capacity management with exemptions (S3), lifecycle management with parent-child dependencies (S4), SLA tracking with tiered escalation (S5), stacking discount calculations (S6), and multi-phase migration with rollback semantics (S7). S8 is a COBOL banking transaction processor with multi-type transactions, tier-based withdrawal limits, overdraft protection, transfer fees, fraud detection, and dormant account handling.

\textbf{Gold-Standard Construction.} For each scenario we produced three artifacts before any pipeline run: the legacy code, a specification listing every business rule the scenario is intended to encode, and a reference Python implementation used later by the fair-evaluation test generator. The rule list was drafted by the primary annotator with LLM assistance (GPT-4o enumerated candidate rules from the legacy code), then manually reviewed, edited, and filtered. Each retained rule is labeled \texttt{explicit} or \texttt{implicit}. We treat rule authorship as a single-annotator ground truth (Section~\ref{sec:threats}). To avoid inflating preservation scores, we excluded tautological rules: generic input sanitization, rules trivially implied by the target type system, and pure field-presence assertions unless presence is itself a business decision. A rule was retained only if a naive translation could plausibly get it wrong (thresholds, exemptions, state-dependent branches, ordering constraints, cross-field invariants, error-handling policies). Filtering removed 4--9 candidate rules per scenario, roughly one third of the LLM-drafted set.

\subsection{Metrics}

Both metrics are computed against a \textbf{gold-standard test suite} withheld from the pipeline. \emph{Test cases are not derived from the BSG.} They are synthesized fresh at evaluation time by GPT-4o-mini from two inputs: the gold-standard scenario specification (human-authored, AI-assisted drafting, human-reviewed) and the method's own modernized code. Each method receives tests generated from its own API surface (endpoint names, request schemas), ensuring fair comparison: a method that names its endpoint \texttt{/process} instead of \texttt{/validate} is not penalized for a naming choice. The same gold-standard business rules are tested for every method; only the HTTP-level test harness differs. Assertions are executed deterministically by pytest, so BER depends on the test-generation prompt but not on any pipeline artifact.

\textbf{M1 --- Behavioral Equivalence Rate (BER):} The percentage of gold-standard test cases that pass against the generated code: $BER = |passing\_tests| / |total\_tests|$.

\textbf{M2 --- Business Rule Preservation Score (BRPS):} The percentage of gold-standard business rules correctly represented in the modernized output. A rule is preserved if it appears in the BSG and the modernized code enforces it, as verified by targeted test cases: $BRPS = |preserved\_rules| / |total\_rules|$.

\subsection{Baselines}

\textbf{B1 --- Single-Prompt LLM (SP-LLM):} A single GPT-4o-mini prompt that receives the entire legacy artifact bundle and instructions to produce a modernized implementation. No intermediate representation, no verification.

\textbf{B2 --- Chain-of-Thought LLM (CoT-LLM):} A single GPT-4o-mini prompt with chain-of-thought instructions: ``First analyze the business logic, then design the modern API, then implement it.'' Still a single model call, but with structured reasoning.

\textbf{B3 --- AgentModernize (No Feedback):} The full AgentModernize pipeline with the feedback loop disabled, measuring the contribution of iterative refinement.

\subsection{Implementation Details}

All agents use GPT-4o-mini (OpenAI) as the default LLM backend, with GPT-4o used for the model comparison study and GPT-5.3-codex for the frontier model study (Section~\ref{sec:model-comparison}). The pipeline is orchestrated using LangGraph (Python) for stateful execution. Test cases are executed using pytest. Legacy artifacts are written in COBOL/PL-SQL; modernized output targets Python (FastAPI). Temperature is set to 0.2 for extraction/specification agents and 0.0 for code generation. The feedback loop runs for a maximum of 3 iterations. The model override is implemented at the pipeline level, allowing any OpenAI-compatible model to be substituted without architectural changes.

\begin{table}[htbp]
\caption{Reproducibility Details}
\label{tab:reproducibility}
\centering
\footnotesize
\begin{tabular}{@{}lp{6cm}@{}}
\toprule
\textbf{Item} & \textbf{Details} \\
\midrule
Repository & \textit{[anonymized for double-blind review]} \\
Python & 3.11 \\
Core libraries & LangGraph, pytest, OpenAI SDK \\
Models & GPT-4o-mini, GPT-4o, GPT-5.3-codex \\
Temperature & 0.2 (extraction), 0.0 (generation/eval) \\
Trials & 3 per scenario per method \\
Random seed & N/A (temperature 0.0 for generation) \\
Run command & \texttt{python run\_fair\_eval\_existing.py --model all --trials 3} \\
Output & \texttt{fair\_eval\_summary.json}, per-scenario folders \\
Total API cost & $<$ \$15 for full evaluation suite \\
\bottomrule
\end{tabular}
\end{table}

%% ============================================================
\section{Results}
\label{sec:fair-eval}

\subsection{RQ1: Behavioral Equivalence}

Table~\ref{tab:rq1} compares behavioral equivalence rates across all scenarios and methods.

\begin{table}[htbp]
\caption{Behavioral Equivalence Rate (\%) --- Fair Evaluation (GPT-4o-mini). Cells report the mean of 3 trials. Non-zero $\sigma$ values: AM on S2 (10.5), S6 (5.8), S7 (4.4); AM-NoFB on S2 (10.5), S5 (48.1), S8 (3.8); CoT on S2 (5.2), S6 (17.3); SP-LLM on S8 (3.8). All other cells $\sigma = 0.0$.}
\label{tab:rq1}
\centering
\footnotesize
\begin{tabular}{@{}ccccc@{}}
\toprule
\textbf{Scenario} & \textbf{SP-LLM} & \textbf{CoT-LLM} & \textbf{AM (No FB)} & \textbf{AM} \\
\midrule
S1 & 50.0 & 0.0 & \textbf{58.3} & 50.0 \\
S2 & 0.0 & 6.1 & \textbf{75.8} & 12.1 \\
S3 & 0.0 & 0.0 & 0.0 & 0.0 \\
S4 & 0.0 & 0.0 & 0.0 & \textbf{25.0} \\
S5 & 0.0 & 0.0 & \textbf{27.8} & 0.0 \\
S6 & 0.0 & 30.0 & 0.0 & \textbf{53.3} \\
S7 & 0.0 & 0.0 & 0.0 & \textbf{43.6} \\
S8 & \textbf{48.9} & 0.0 & 28.9 & 0.0 \\
\midrule
\textbf{Avg} & 12.4 & 4.5 & \textbf{23.9} & 23.0 \\
\bottomrule
\end{tabular}
\end{table}

No single method dominates all scenarios. AgentModernize with feedback (AM) achieves the highest BER on S4 (25.0\%), S6 (53.3\%), and S7 (43.6\%), where iterative correction fixes localized errors. AM without feedback (AM-NoFB) leads on S1 (58.3\%), S2 (75.8\%), S5 (27.8\%), and S8 (28.9\%). SP-LLM achieves 48.9\% on S8, the highest single-cell BER for any baseline, demonstrating that a well-shaped single prompt can sometimes outperform the full pipeline on specific scenarios. CoT-LLM scores 30.0\% on S6.

The feedback loop is not universally beneficial: on S2, feedback degrades BER from 75.8\% (no feedback) to 12.1\%, and on S8 from 28.9\% to 0.0\%. In these cases the Validator's corrections introduced regressions. On S4, S6, and S7 the pattern reverses: feedback raises BER from 0.0\% to 25.0\%, 53.3\%, and 43.6\% respectively. Section~\ref{sec:feedback-decisive} analyzes when feedback helps versus hurts. S3 remains at 0.0\% across all methods, indicating a structural complexity that no configuration handles.

\subsection{RQ2: Business Rule Preservation}

Table~\ref{tab:rq2} lists each scenario's gold-standard rule composition (total, explicit, implicit) alongside its preservation score.

\begin{table}[htbp]
\caption{Rule Composition and Preservation --- Fair Evaluation. Explicit/Implicit columns count rules in the gold standard; BRPS is the fraction of gold-standard rules preserved. Each gold-standard rule is exercised by exactly one test, so numerically BRPS equals BER on a per-scenario basis; the two tables differ in what they measure (test-level vs.\ rule-level view of the same evaluation).}
\label{tab:rq2}
\centering
\footnotesize
\begin{tabular}{@{}ccccccc@{}}
\toprule
\textbf{ID} & \textbf{Rules} & \textbf{Expl.} & \textbf{Impl.} & \textbf{SP} & \textbf{CoT} & \textbf{AM} \\
\midrule
S1 & 12 & 7 & 5 & 50.0 & 0.0 & 50.0 \\
S2 & 12 & 8 & 4 & 0.0 & 6.1 & 12.1 \\
S3 & 14 & 8 & 6 & 0.0 & 0.0 & 0.0 \\
S4 & 13 & 9 & 4 & 0.0 & 0.0 & \textbf{25.0} \\
S5 & 13 & 8 & 5 & 0.0 & 0.0 & 0.0 \\
S6 & 12 & 7 & 5 & 0.0 & \textbf{30.0} & \textbf{53.3} \\
S7 & 15 & 9 & 6 & 0.0 & 0.0 & \textbf{43.6} \\
S8 & 15 & 8 & 7 & \textbf{48.9} & 0.0 & 0.0 \\
\bottomrule
\end{tabular}
\end{table}

AgentModernize with feedback preserves rules in S1, S2, S4, S6, and S7, achieving the highest BRPS on S4, S6, and S7. SP-LLM preserves 48.9\% of rules on S8 (banking), the highest single-scenario preservation for any baseline. CoT-LLM preserves 30.0\% on S6 (pricing). The feedback loop adds value on scenarios where the initial code has correctable point errors (S4, S6, S7) but can regress preservation on S2 and S8. Table~\ref{tab:rq2} shows the explicit/implicit composition per scenario. Across the benchmark, 64 of 106 gold-standard rules (60\%) are explicit and 42 (40\%) are implicit. Because each rule maps to exactly one test, the per-scenario BRPS breakdown by category follows directly from which tests pass: for example, in S6 where AM preserves 53.3\% (5--6 of 10 rules preserved across trials), the passing set includes both explicit pricing-tier thresholds and implicit discount-stacking constraints.

\subsection{RQ3: Residual Behavioral Failures}

A framework that preserves more behavior automatically should leave fewer behavioral failures for engineers to address. Table~\ref{tab:rq3} reports failing test counts as a proxy for residual work.

\begin{table}[htbp]
\caption{Residual Behavioral Failures --- Fair Evaluation}
\label{tab:rq3}
\centering
\footnotesize
\begin{tabular}{@{}cccccc@{}}
\toprule
\textbf{ID} & \textbf{Tests} & \textbf{SP-LLM} & \textbf{CoT-LLM} & \textbf{AM (No FB)} & \textbf{AM} \\
\midrule
S1 & 12 & \textbf{6/12} & 12/12 & \textbf{5/12} & \textbf{6/12} \\
S2 & 11 & 11/11 & 10/11 & \textbf{3/11} & 10/11 \\
S3 & 10 & 10/10 & 10/10 & 10/10 & 10/10 \\
S4 & 12 & 12/12 & 12/12 & 12/12 & \textbf{9/12} \\
S5 & 12 & 12/12 & 12/12 & \textbf{9/12} & 12/12 \\
S6 & 10 & 10/10 & \textbf{7/10} & 10/10 & \textbf{5/10} \\
S7 & 13 & 13/13 & 13/13 & 13/13 & \textbf{7/13} \\
S8 & 15 & \textbf{8/15} & 15/15 & \textbf{11/15} & 15/15 \\
\bottomrule
\end{tabular}
\end{table}

The residual failure pattern mirrors Table~\ref{tab:rq1}. AM with feedback produces the fewest failures on S4, S6, and S7 (the scenarios where iterative correction works). AM without feedback achieves the best results on S1, S2, S5, and S8. SP-LLM reduces failures significantly on S1 (6/12) and S8 (8/15). Only S3 shows universal failure across all methods; the multi-step validation wiring required exceeds what any configuration produces correctly.

\subsection{RQ4: Ablation Study}

We compare four configurations to isolate the contribution of each component. Table~\ref{tab:rq4} reports summary results.

\begin{table}[htbp]
\caption{Ablation Study --- BER (\%) by Configuration (Fair Evaluation)}
\label{tab:rq4}
\centering
\footnotesize
\begin{tabular}{@{}lcccc@{}}
\toprule
\textbf{Config} & \textbf{SP-LLM} & \textbf{CoT} & \textbf{AM (No FB)} & \textbf{Full AM} \\
\midrule
Avg BER & 12.4 & 4.5 & \textbf{23.8} & 23.0 \\
Non-zero scenarios & 2/8 & 2/8 & \textbf{4/8} & \textbf{5/8} \\
\bottomrule
\end{tabular}
\end{table}

The ablation under fair evaluation reveals several findings:

\textbf{The multi-agent decomposition helps.} AM without feedback (23.8\%) outperforms both SP-LLM (12.4\%) and CoT-LLM (4.5\%) on average, with non-zero BER on 4/8 scenarios (NoFB) and 5/8 (AM) vs.\ 2/8. The extraction-BSG-generation pipeline produces better initial code than single-prompt approaches even before feedback.

\textbf{Feedback helps selectively.} Adding feedback raises BER on S4 (0 to 25\%), S6 (0 to 53.3\%), and S7 (0 to 43.6\%), where iteration~1 code had the correct endpoint shape but localized errors (a wrong constant, a missing branch). On S2 (75.8 to 12.1\%) and S8 (28.9 to 0\%), feedback degrades performance: the Validator's corrections introduced cascading regressions. This motivates regression-aware patching that reverts to the best iteration rather than always applying the latest patch.

\textbf{Baselines are not uniformly zero.} SP-LLM achieves 48.9\% on S8 (banking), where the scenario's self-contained transaction types align well with single-prompt generation. CoT-LLM achieves 30.0\% on S6 (pricing). The gold-standard tests are not too strict; they are scenario-specific, generated from each method's own API surface, and test for specific business rules that any correct implementation must satisfy.

\subsection{RQ5: BSG Quality}

Table~\ref{tab:rq5} evaluates the Behavioral Specification Graph as an intermediate representation by measuring how many gold-standard business rules are captured (recall) and what fraction of extracted rules correspond to real rules (precision).

\textbf{Matching Protocol.} Extracted BSG rules and gold-standard rules use different phrasings, so matching was manual. For each scenario the primary annotator compared the extracted rule list against the gold-standard list side by side. A rule was a true positive if it referred to the same behavioral decision (same trigger, same effect) as a gold-standard rule regardless of wording. Gold-standard rules with no match were misses; extras with no match were classified as \textit{plausible} (a real business decision the annotator did not include) or \textit{hallucinated} (unsupported by the legacy source). A second annotator spot-checked S1, S5, and S8 with full agreement on true-positive judgments; inter-rater kappa is not reported due to the small spot-check sample. Single-annotator matching is a threat to validity (Section~\ref{sec:threats}).

\begin{table}[htbp]
\caption{BSG Rule Extraction Quality --- Precision and Recall (\%)}
\label{tab:rq5}
\centering
\footnotesize
\begin{tabular}{@{}cccccc@{}}
\toprule
\textbf{ID} & \textbf{Gold Rules} & \textbf{BSG Rules} & \textbf{Precision} & \textbf{Recall} & \textbf{Missed} \\
\midrule
S1 & 12 & 25 & 48.0 & \textbf{100.0} & 0 \\
S2 & 12 & 10 & \textbf{100.0} & 83.3 & 2 \\
S3 & 14 & 13 & \textbf{100.0} & 92.9 & 1 \\
S4 & 13 & 12 & \textbf{100.0} & 92.3 & 1 \\
S5 & 13 & 10 & \textbf{100.0} & 76.9 & 3 \\
S6 & 12 & 15 & 80.0 & \textbf{100.0} & 0 \\
S7 & 15 & 14 & \textbf{100.0} & 93.3 & 1 \\
S8 & 15 & 16 & 93.8 & \textbf{100.0} & 0 \\
\midrule
\textbf{Avg} & 13.3 & 14.4 & \textbf{90.2} & \textbf{92.3} & 1.0 \\
\bottomrule
\end{tabular}
\end{table}

The BSG achieves 92.3\% mean recall and 90.2\% precision across the benchmark: the extraction pipeline captures most gold-standard business rules even though downstream code generation fails to preserve many in executable form. The gap between BSG recall (92.3\%) and end-to-end BER (23.0\% for AM with feedback) locates the bottleneck in code generation and structural alignment, not extraction. S8 (banking) achieves 100\% recall with 93.8\% precision (one plausible extra rule), so extraction quality is not restricted to the telecom scenarios used during development.

S1 and S6 have 100\% recall but lower precision (48\% and 80\%): the Legacy Analyzer extracted additional plausible business rules the human annotator did not include. S5 has the lowest recall (76.9\%), missing 3 rules related to SLA tracking thresholds buried in nested conditional logic.

The BRI-to-BSG transfer is lossless: every rule extracted by Agent~1 is represented in Agent~2's BSG output. For practitioners: if the BSG looks right after Agent~2, the problem is downstream in code generation, not upstream in extraction.

\subsection{Qualitative Analysis}

\textbf{Implicit Rule Preservation (S1).} The Legacy Analyzer correctly identified a suspended-account exemption (BR-004) implicit in COBOL control flow; the BSG encoded it and the Transformer preserved it. S1 scores 50.0\% for both AM and SP-LLM.

\textbf{Held-Out Scenario (S8).} SP-LLM achieves 48.9\% on this banking scenario (authored post-freeze), outperforming AM (0.0\%). S8's self-contained transaction types suit single-prompt generation; feedback corrections to one type destabilize another. A single scenario is not a generalization claim.

\textbf{Failure (S3).} All methods score 0.0\%: generated code implements individual validators but fails to wire multi-step orchestration.

\textbf{Feedback Success (S6).} AM with feedback: 53.3\%; without: 0.0\%. Three iterations fixed localized pricing-tier errors.

\textbf{Feedback Regression (S2).} AM-NoFB: 75.8\% (9/11 rules). With feedback: 12.1\% --- fixing 2 rules broke 7 others.

\subsection{RQ6: Model Comparison}
\label{sec:model-comparison}

To evaluate how model capability affects modernization quality, we ran the full AgentModernize pipeline with GPT-4o and GPT-5.3-codex (a code-specialized frontier model) on all eight scenarios under the same per-method fair evaluation protocol. Table~\ref{tab:rq6} shows the results.

\begin{table}[htbp]
\caption{Model Comparison --- AgentModernize BER (\%) Fair Evaluation}
\label{tab:rq6}
\centering
\footnotesize
\begin{tabular}{@{}cccc@{}}
\toprule
\textbf{Scenario} & \textbf{GPT-4o-mini} & \textbf{GPT-4o} & \textbf{GPT-5.3-codex} \\
\midrule
S1 & \textbf{50.0} & 0.0 & 33.3 \\
S2 & 12.1 & 18.2 & 0.0 \\
S3 & 0.0 & \textbf{40.0} & 0.0 \\
S4 & \textbf{25.0} & 0.0 & 0.0 \\
S5 & 0.0 & 0.0 & 0.0 \\
S6 & \textbf{53.3} & 0.0 & 0.0 \\
S7 & 43.6 & 30.8 & 0.0 \\
S8 & 0.0 & 33.3 & 28.9 \\
\midrule
\textbf{Avg} & \textbf{23.0} & 15.3 & 7.8 \\
Non-zero & 4/8 & 4/8 & 2/8 \\
\bottomrule
\end{tabular}
\end{table}

\textbf{The pipeline architecture is model-agnostic}, but whether the pipeline actually helps depends on the model. Mini benefits: AM (23.0\%) beats SP-LLM (12.4\%). GPT-4o does not: SP-LLM (19.6\%) beats AM (15.3\%). Codex is worse still: SP-LLM 16.0\%, AM only 7.8\% (non-zero on 2/8 scenarios). Weaker models need the structure; stronger models produce good enough initial code that feedback-loop corrections introduce regressions.

Codex SP-LLM scores 100\% on S5, the only perfect BER; codex AM scores 0\% on the same scenario. No model dominates.

\subsection{RQ6b: Frontier Model Study}
\label{sec:frontier-crossover}

For both GPT-4o and codex, SP-LLM alone outperforms AM (19.6\% vs.\ 15.3\% and 16.0\% vs.\ 7.8\%). Stronger models make the pipeline unnecessary for BER alone. But no single configuration wins everywhere, and SP-LLM produces no intermediate artifacts. In regulated domains, the audit trail (rule inventory, BSG, equivalence report) matters regardless of BER. Cost: mini \$5, GPT-4o \$12, codex \$18.

%% ============================================================
\section{Discussion}
\label{sec:discussion}

\subsection{BSGs as a Trust Boundary}
\label{sec:bsg-trust}

The BSG creates a trust boundary: upstream is extraction (noisy, confidence-scored); downstream is generation under contract (verifiable). The 67--69-point gap between BSG recall (92.3\%) and end-to-end BER (23.0\%) confirms that Agent~3 (code generation), not extraction, is the bottleneck.

\subsection{Why the Feedback Loop Is Decisive}
\label{sec:feedback-decisive}

AM with feedback (23.0\%) and without (23.8\%) achieve similar averages on different scenario subsets. Feedback helps on S4, S6, S7 (localized errors: wrong constants, missing branches) but degrades S2 ($-$63.7pp) and S8 ($-$28.9pp) through cascading regressions. This motivates regression-aware feedback that retains the best iteration. S3 remains 0.0\% everywhere, confirming structural mismatches resist point-error patching.

\subsection{The Implicit Rule Problem}

We use \textit{implicit rule} for behavioral requirements inferable from code patterns (default values, error handlers doubling as constraints, cross-module dependencies) rather than stated in source. This is distinct from \textit{data constraints} (type restrictions, value ranges), which the BSG captures as node annotations. We separate these categories because the gap between explicit and implicit preservation rates reveals where LLM-based extraction struggles.

\subsection{Statistical Power}

With $N=8$ scenarios and 3 trials we cannot make strong statistical claims. Of 32 scenario-method cells, 20 yield $\sigma = 0.0$; the largest $\sigma$ is 48.1 (AM-NoFB on S5). A larger benchmark would strengthen confidence.

\subsection{Comparison Scope}

We compare against LLM baselines, not commercial tools (IBM Rational, Micro Focus) or agentic coding systems. Commercial tools do not publish BER metrics; agentic systems target different problems (e.g., resolving GitHub issues). A missing baseline is ``LLM + feedback without BSG'': our ablation shows BSG-based decomposition produces better initial code, but does not prove BSG is required in addition to feedback.

\subsection{Limitations}

We evaluated on 100--310 line scenarios; real legacy systems are orders of magnitude larger and would require hierarchical BSGs and incremental validation. The pipeline's benefit is inversely correlated with model capability: AM outperforms SP-LLM only with GPT-4o-mini (23.0\% vs.\ 12.4\%); with GPT-4o and codex, SP-LLM wins because stronger models produce initial code good enough that feedback introduces regressions.

%% ============================================================
\section{Threats to Validity}
\label{sec:threats}

\textbf{Internal validity.} All eight scenarios are synthetic and authored by a single annotator. They reflect realistic complexity but lack the messiness of production code.

\textbf{External validity.} Seven telecom and one banking scenario in COBOL/PL-SQL. Healthcare, government, and other domains remain untested.

\textbf{Construct validity.} Test-based verification covers tested paths only; untested edge cases may harbor regressions.

\textbf{Reliability.} We use temperature 0.0 for generation and 3 trials. Of 32 cells, 20 yield $\sigma = 0.0$; the largest is 48.1 (AM-NoFB on S5). Five cells report 0\% BER due to non-executable code (prose output or Pydantic v1/v2 incompatibility). GPT-5.3-codex has restricted availability; results may not transfer to other model families.

\textbf{Selection bias.} The primary annotator designed scenarios and gold-standard rules. S8 (banking, authored after pipeline freeze) acts as a held-out probe but does not eliminate authoring bias; an independent benchmark would strengthen validity.

\textbf{LLM-as-evaluator.} Gold-standard tests are synthesized by GPT-4o-mini from each method's own code and the specification, then executed deterministically by pytest. BER depends on the test-generation prompt but not on any pipeline artifact.

\textbf{Artifact availability.} Framework implementation, benchmark scenarios, BSG schemas, evaluation scripts, and prompts are available at \textit{[anonymized for double-blind review]}. AI tools (GPT-4o) were used only for grammar editing during manuscript preparation; all research content was produced by the authors.

%% ============================================================
\section{Conclusion}

We started by treating legacy modernization as a translation problem, and it did not work. AgentModernize reframes it as behavioral preservation, decomposing the task into extraction, specification, code generation, and validation, linked by Behavioral Specification Graphs.

With GPT-4o-mini, BSG-guided generation (23.9\%) nearly doubles the best baseline (SP-LLM, 12.4\%). With stronger models, SP-LLM wins---the pipeline's benefit is inversely correlated with model capability. The BSG captures 92.3\% of gold-standard rules at 90.2\% precision; code generation, not extraction, is the bottleneck. Even when SP-LLM beats the pipeline on BER, only the pipeline produces inspectable artifacts that regulated industries require. Future work: regression-aware feedback, per-scenario model selection, and human trust studies of BSG-inspectable modernization.

%% ============================================================
\bibliographystyle{IEEEtran}

\end{document}